\documentclass[aps,twocolumn,amsmath,amssymb,floatfix]{revtex4-1}

\usepackage{graphicx}
\usepackage{dcolumn}
\usepackage{bm}
\usepackage{epsfig,amsmath,amssymb,color,calc}

\newcommand{\ab}{{\it ab initio }}

\newcommand{\Q}[1]{Q$^{(#1)}$}

\newcommand{\boric}{{B$_2$O$_3$}}
\newcommand{\silica}{{SiO$_2$}}
\newcommand{\soda}{{Na$_2$O}}
\newcommand{\B}[1]{$^{[#1]}$B}

\newcommand{\biii}{$^{[3]}$B\,}
\newcommand{\biv}{$^{[4]}$B\,}
\newcommand{\bnmr}{$^{11}$B}
\newcommand{\si}{$^{29}$Si}
\newcommand{\ox}{$^{17}$O}

\newcommand{\bo}{BO}
\newcommand{\nbo}{NBO\, }
\newcommand{\tbo}{TBO\, }

\begin{document}

\title{FIRST PRINCIPLES STUDY OF A SODIUM BOROSILICATE GLASS-FORMER I: THE LIQUID STATE}

\author{Laurent Pedesseau}
 \altaffiliation{present address: Universit\'e Europ\'enne de Bretagne,
INSA, FOTON, UMR 6082, 35708 Rennes}
\author{Simona Ispas}
 \email[]{simona.ispas@univ-montp2.fr}
 \author{Walter Kob}
 \email[]{walter.kob@univ-montp2.fr}
\affiliation{Laboratoire Charles Coulomb, UMR
5221, Universit\'e Montpellier 2 and CNRS, 34095 Montpellier, France }

\begin{abstract}

We use {\it ab initio} simulations to study the static and
dynamic properties of a sodium borosilicate liquid with composition
3Na$_2$O-B$_2$O$_3$-6SiO$_2$, i.e. a system that is the basis of many glass-forming
materials. In particular we focus on the question how boron is embedded
into the local structure of the silicate network liquid. From the partial
structure factors we conclude that there is a weak   nanoscale phase
separation between silicon and boron and that the sodium atoms form
channel-like structures as they have been found in previous studies
of sodo-silicate glass-formers. Our results for the X-ray and neutron
structure factor show that this feature is basically unnoticeable in
the former but should be visible in the latter as a small peak at
small wave-vectors. At high temperatures we find a high concentration
of three-fold coordinated boron atoms which decreases rapidly with
decreasing $T$, whereas the number of four-fold coordinated boron atoms
increases. Therefore we conclude that at the experimental glass transition
temperature most boron atoms will be four-fold coordinated. We show
that the transformation of $^{[3]}$B into $^{[4]}$B with decreasing $T$
is not just related to the diminution of non-bridging oxygen atoms as
claimed in previous studies, but to a restructuration of the silicate
matrix. The diffusion constants of the various elements show an Arrhenius
behavior and we find that the one for boron has the same value as the
one of oxygen and is significantly larger than the one of silicon. This
shows that these two network formers have rather different dynamical
properties, a result that is also confirmed from the time dependence of
the van Hove functions. Finally we show that the coherent intermediate
scattering function for the sodium atoms is very different from the
incoherent one and that it tracks the one of the matrix atoms.

\end{abstract}


\maketitle

\section{\label{sec:intro}Introduction}

Borosilicate glasses have many remarkable properties such as a
low thermal expansion coefficient, weak electrical conductivity,
high resistance to thermal shocks, and good stability regarding
corrosion~\cite{Varshneya_book,Greaves20071}. Thanks to these
features, these glasses have widespread applications going from
every-day kitchenware to laboratory glassware, from insulating
materials to those used for immobilisation of the nuclear waste
\cite{Trotignon1992228,Bardez2006272,McKeown201013,Delaye20112763}. The
borosilicates that are of technological interest contain, apart from
silicon and boron oxides, also a certain amount of network modifiers
such as alkali and alkaline-earth oxides, as well as network formers
such as Al or P. It is the resulting complex structure that is believed
to give these glasses their remarkable properties~\cite{Varshneya_book}
and hence understanding this structure poses also an interesting challenge
for fundamental science.

One possibility to gain insight how the distinctive features of
these glasses are related to their composition is to study a series
of simple compositions, e.g. the ternary alkali borosilicates
M$_2$O-B$_2$O$_3$-SiO$_2$, with M=Li, Na, K. Despite the apparent
simplicity of these alkali borosilicates, they present non trivial
physical and chemical behavior under a change of composition, temperature,
pressure, or irradiation and hence such comparative studies have allowed
to understand some of the connections between structure and properties
\cite{Sen199829,Sen199984,Martens2000167,Du2003239,Chen2004104,Du2004196,Stebbins200880,Manara20092528,Wu20113944,Angeli2012054110}.
These kind of studies have demonstrated the need to obtain quantitative
information on the factors responsible for the properties that make  these
materials so important for glass technology, and in order to achieve
this goal it has become mandatory to understand  their structure on the
atomistic scale. Hence, one needs an answer to the simple but probably the
most fundamental question: How does boron modify the structure/integrate
into the silica network? Answering this question will help to design
new compositions that are energy- and environmentally-friendly and hence
needed to make progress in the field.

In the late seventies and early eighties of the 20$^{th}$ century, several
studies have been carried out for the ternary composition containing
sodium oxide, i.e.~\soda-\boric-\silica. Many of these studies were
done by Yun, Bray, Dell and co-workers, using solid-state nuclear
magnetic resonance spectroscopy (NMR) \cite{Yun01,Yun02,Dell19831}
of \bnmr.  Based on these experiments, a structural model has been
proposed (called hereafter YBD) in order to describe the evolution
of the structure as cations (Na) atoms are added to the melt and the
mechanism of creation of non-bridging oxygens. This evolution is usually
parameterized in terms of two ratios $K=[$\silica $]/[$\boric $]$  and $
R=[$\soda$]/[$\boric$]$ ($[.]$ indicate mol\%).  Using only the quantities
$K$ and $R$, the YBD model assumes that the borosilicate glasses contain
several larger structural units like diborate, pyroborate, boroxol rings,
reedmergnerite, danburite, etc...\cite{Varshneya_book}.  These units (also
called  supra-structural units) are in turn composed  of basic units,
such as four-coordinated silicon, three- and four-coordinated borons,
and within the model one divides the \soda-\boric-\silica~ternary diagram
into four compositional regions.  For   every compositional domain,
the YBD model predicts the fraction of three- and four-fold coordinated
borons (\biii and \biv) among total boron concentration as well as the
fraction of  bridging oxygens.

Following up these early \bnmr \, NMR studies, other techniques were
employed to explore the distribution of the structural groups and
the mixing of silicate and borate units: combined Raman and \bnmr \,
NMR  \cite{Bunker199030}, X-ray absorption near-edge structure (XANES)
\cite{Fleet1999233}, infra-red (IR) \cite{Kamitsos199431}, as well as
\si, \ox\,  and \bnmr\,  NMR \cite{Martens2000167}.  Further significant
progress in understanding how silicate structural units mix with the
borate \B3 and \B4 units has been made during the last two decades,
and this was the direct consequence of the technical advances in
solid-state NMR  experiments, with the emergence of high-resolution
magic-angle spinning (MAS)  and especially  multiple-quantum magic-angle
spinning NMR techniques.  For example Stebbins and co-workers
have reported results on the concentration of \B4~units, and also
made assignments of the various oxygen sites, namely the Si-O-Si,
B-O-B, Si-O-B and [Si,B]-O-Na linkages \cite{Sen199829,Wang1998286,
Wang19991519,Lee200112583, Du200310063}.  The evolution of the  \biv
fraction has in fact attracted much interest, and various experimental
studies have predicted that this concentration decreases if the quench
rate increases \cite{Sen199829,Wu20102097,Angeli2012054110}.

Experiments show that borosilicate liquids and glasses present non-linear
changes of their macroscopic properties with varying composition,
temperature and pressure, and present days computer simulations
can provide  valuable atomic-scale information on both structure and
dynamics.  For the particular case of oxide glasses and liquids, atomistic
simulations have become a well established tool for getting  insight into
the processes taking place at the microscopic level, known to control
then the  macroscopic properties \cite{Binder2005713,Tilocca20091003}.
While the physics and chemistry of pure liquid and glassy \silica\,
\cite{Binder2005713,Pedone200920773,Giacomazzi2009064202} and \boric\,
\cite{Takada19958659,Takada19958693,Umari2005137401,Huang2006224107,Ferlat2008065504,
Ohmura2008224206,Ohmura2009020202,Ohmura2010014208}, have been intensively
studied using simulations, there are so far only few numerical studies dedicated
to ternary sodium borosilicate.  The majority of those  studies rely on
the use of effective potentials, i.e.~a classical molecular dynamics (MD)
approach.  The very first one was reported thirty years ago by Soules and
Varshenya \cite{Soules1981145}, who used a pair potential and primarily
studied the boron coordination changes when the composition changes.
Then a decade ago, Gou et al. \cite{Gou2001539} examined again the
structure of some sodium borosilicate glasses using a three-body effective
potential, and concluded that there was a tendency for the borate network
to separate from the silicate part together with an association of sodium
with the former. Very recently, Kieu et al. \cite{Kieu20113313}  have
proposed a class of pair-potentials dedicated to this ternary system,
and included a dependence between fitting parameters and composition in
order to better reproduce the structural and mechanical properties over
a wide compositional range.  This approach allowed to reproduce certain
aspects of the so-called boron anomaly~\cite{Varshneya_book}.

Although simulations with effective potentials can certainly give
valuable insight into the structural and dynamical properties
of glass-forming systems, it is far from evident that they give
quantitatively good results for multi-component systems since usually
no reliable potentials are available.  Hence for such systems it
is preferable to use {\it ab initio} simulations based on density
functional theory since these can handle also more complex local
atomic environments. However, since often the concentration of
one species is rather low, and {\it ab initio} simulations become
computationally very expensive if the system size is large (say several
hundred atoms), there exist so far relatively few studies that used this
approach~\cite{Ganster200410172,Tilocca20061950,Du2006114702,Ispas2010,Christie20112038}.
For borosilicate glasses, the only \ab investigation so far reported in
the literature, is a study of  the structural and energetic effects of
sodium substitution by hydronium ions  \cite{Geneste20063147}.

The goal of the present study, as well as of the companion paper
\cite{Pedesseau-nbs2}, is therefore to use {\it ab initio} simulations
to obtain insight into the structure and dynamics of a sodium-rich
borosilicate liquid and glass as a function of temperature.
The composition of our system is 3~Na$_2$O-B$_2$O$_3$-6~SiO$_2$
(called NBS hereafter), and is similar to the composition used in
glass wool. Within the YBD terminology, the present composition
corresponds to $R=3$ and $K=6$. From an experimental point of view,
there exist quite a few studies on the present composition in that NMR,
Raman, XANES, and XPS experiments have been reported more than 10
years ago \cite{Dell19831,Wang1998286,Bunker199030,Fleet1999233,
Miura20011}, and very recently a neutron diffraction study has been
done~\cite{Michel2013169}.   These experimental studies have focused
on the structural features of the glassy state and in the following
we will compare these results with ours. In addition we mention that
compositions quite close to  our NBS system  have been investigated by IR
\cite{Kamitsos199431} and  NMR \cite{Martens2000167} experiments, while
Yamashita et al. \cite{Yamashita2001535} have studied its thermodynamic
properties, and, taking into account the contributions of the structural
units, have built a model for computing the heat capacity.  We thus will
discuss our results also with respect to these studies.

The paper is organized as follows: In the next section we present
the details of the simulations. Section~\ref{sec:liqstruct}
we will discuss the structural properties of the liquid, and in
Sec.~\ref{sec:DP} the dynamical ones. Finally we summarize the results in
Sec.~\ref{sec:conclu}. Whereas the present paper focuses on the liquid,
the accompanying paper, to which we will refer to as Part~II, is devoted to
the structural, electronic and vibrational properties of the glass.

\section{\label{sec:method}Simulation details}

The {\it ab initio} MD simulations were done using the Vienna {\it ab
initio} package (VASP) \cite{vasp_code01,vasp_code02}.  The system we
have considered has the composition 3 Na$_2$O-B$_2$O$_3$- 6 SiO$_2$
and we have used a cubic box containing 320 atoms (60 silicon, 180 oxygen,
60 sodium and 20 boron atoms) and periodic boundary conditions. The edge
length of the box has been fixed to 15.97 \AA, which corresponds to
the experimental NBS mass density of 2.51 g/cm$^3$ \cite{OMazurin_book}.

The electronic structure has been calculated by means of the
Kohn-Sham (KS) formulation of the Density Functional Theory
(DFT) \cite{Kohn1965,RMartin_book} using the generalized gradient
approximation (GGA) and the PBEsol functional \cite{GGAPBE,PBEsol}.
The choice of the recently proposed PBEsol functional \cite{PBEsol}
to describe the electronic exchange and correlation is motivated by
the fact that for equilibrium structures and vibrational spectra
of extended systems it often gives better results than other GGA
functionals~\cite{Demichelis2010406}. The KS orbitals have been expanded
in a plane wave basis at the $\Gamma$-point of the supercell of the
systems and the electron-ion interaction has been described within the
projector-augmented-wave formalism  \cite{PAW,PAW-vasp}. The plane-wave
basis set contained components with energies up to 600 eV.

In order to solve the KS equations, we have used the residual minimization
method-direct inversion in the iterative space~\cite{vasp_code01}, and the
electronic convergence criterion was fixed at $5\cdot 10^{-7}$ eV.  For
the {\it ab initio} MD simulations, the time step for the motion of the
ions was chosen to be 1~fs and a Nos\'{e} thermostat \cite{Nose1984255}
was applied to control the temperature in the canonical ensemble (NVT). To
determine the vibrational properties of the glass we have cooled the
sample to zero temperature and then determined the local minimum of the
potential energy. This structural relaxation was stopped once the $x,\,
y,\, z -$ components of the forces acting on each atom were inferior
than $10^{-3}$~eV/\AA.

The results presented in the next sections have been obtained by averaging
over 2 independent samples.  To generate an initial configuration, we
have used a random arrangement of atoms placed in the simulation box.
Subsequently we started the {\it ab initio} MD simulations within the
$NVT$ ensemble at 4500~K.  After equilibration at this temperature,
we performed $NVT$ simulations at 4 lower temperatures: 3700~K, 3000~K,
2500~K, and 2200~K. For the three highest $T$'s we discarded the first
0.5~ps from the total length of the runs before we started to measure
the observables of interest, whereas for the two lowest temperatures
we removed the first 1.5~ps. The lengths of the trajectories considered
in the following for studying the structural and dynamic properties of
the liquid were 2~ps for 4500~K, 2.5~ps for 3700~K, 7~ps for 3000~K,
20~ps for 2500~K, and 30~ps for 2200~K.  We note that, except for
the lowest temperature, we stopped the NVT simulations once the mean
squared displacement (MSD) of the slowest element -i.e.~silicon -
reached $\approx$  10~\AA$^2$, which we considered as sufficient to
assure that all the species have reached the diffusive regime.  Due to
the computational cost, we stopped the simulation at 2200~K before
this criterion had been fulfilled. At this $T$ the MSD of silicon
atoms reaches $\approx $ 5.2 \AA$^2$ (see subsec. \ref{sec:MSD} and
Fig. \ref{fig:fig10-msd}d). At each one of these temperatures, we have
computed the pressure of the liquid and we found the following values (for
decreasing $T$): 3.0 GPa, 2.9 GPa, 2.2 GPa, 2.0 GPa and 1.2 GPa. A graph
of these pressures as a function of $T$ shows that for $T\approx$~750~K
the pressure vanishes, i.e. at a temperature which is close to the
experimental value of the glass transition temperature. Therefore we can
conclude that our simulation is indeed able to predict the experimental
value for the density of the glass.

In order to study the structural properties of the glass at room
temperature, we have generated 6 samples that had different thermal
histories due to the variation of the quench rate as well as of the
starting temperature of the quench. More precisely, we have generated
these samples using a two-steps procedure:  Four samples were obtained
by firstly quenching equilibrium configurations from 3000~K to 2000~K,
using a quench rate of $2\times 10^{14}$ K.s$^{-1}$, and subsequently with
a higher rate of $1.7\times 10^{15}$ K.s$^{-1}$ from 2000~K to 300~K.
For the two other samples we used configurations at 2200~K and quenched
them to 1200~K, and then followed the second faster quench down to 300~K,
using a rate of $9\times 10^{14}$ K.s$^{-1}$.  At 300~K, we annealed the
samples for 2~ps using the NVT ensemble. For two samples this annealing
was followed by a run in the NVE ensemble of duration 8~ps and 15~ps,
respectively.  We have found that within statistical fluctuations all
these samples had the same structural properties and therefore we have
averaged these properties over all six samples.  The mean pressure of
these glassy sample was around $-0.04$GPa, which shows that despite the
fast quench rate, we recover the experimental pressure.

\section{\label{sec:liqstruct}Structure}

In this section, we will present and discuss the static  properties
of our NBS liquid samples at the various temperatures considered. In
addition we will also present the corresponding properties of the glass,
even if these will be discussed only in Part II.

\subsection{\label{sec:PDFliq}Radial Pair Distribution Functions}

\begin{figure}
\includegraphics[width=0.43\textwidth]{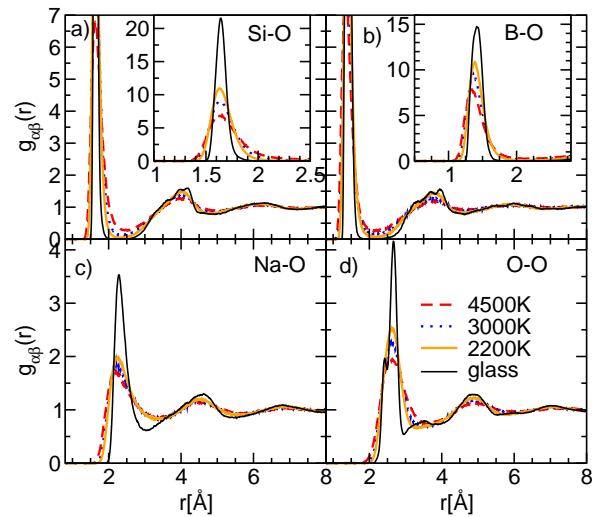}%
\caption{\label{fig:fig1-gdr1} Pair distribution functions for X-O pairs,
(X = Si, O, Na, B)  plotted for the liquid at three temperatures
and the glass state at 300 K. The insets in the upper panels show the
first peaks of the Si-O  and B-O PDFs, respectively.}
\end{figure}

\begin{figure}
\includegraphics[width=0.43\textwidth]{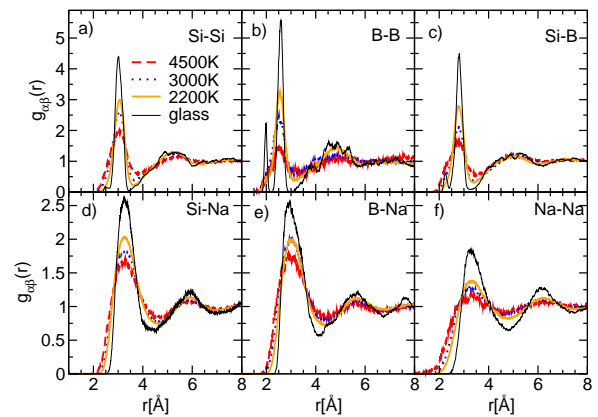}%
\caption{\label{fig:fig2-gdr2} Pair distribution functions $g_{\alpha
\beta} (r)$ for $\alpha, \, \beta =$  Si, Na, B, plotted for the liquid
at three temperatures and the glass state at 300 K. }
\end{figure}

In Figs.~\ref{fig:fig1-gdr1} and ~\ref{fig:fig2-gdr2}, we show
the  partial  pair distribution functions (PDF)  $g_{\alpha\beta}
(r)$ for $\alpha,\, \beta= \mathrm{Si, O, B, Na}$, defined
by~\cite{BinderKob_book}:

\begin{equation}
g_{\alpha \beta}(r ) = \frac{V}{N_\alpha (N_\beta - \delta_{\alpha 
\beta})}
\left \langle \sum_{i=1}^{N_\alpha} \sum_{j=1}^{N_\beta} 
\frac{1}{4\pi r^2} \delta(r- |\vec r_i -\vec r_j|) \right \rangle
\quad,
\end{equation}

\noindent
where $\langle \cdot\rangle$ represents the thermal average, $V$ is the
volume of the simulation box, $N_\alpha$ is the number of particles of
species $\alpha$, and $\delta_{\alpha\beta}$ is the Kronecker delta. We
note that Figs.~\ref{fig:fig1-gdr1} and ~\ref{fig:fig2-gdr2} as well as
all figures discussed in this subsection shows simulation data for both
liquid and  glass at 300 K, but the glass features will be discussed in
Ref.~\cite{Pedesseau-nbs2}. Also we mention that, for the sake of clarity,
we show only the functions for 3 of the 5 temperatures we simulated.

To start, we point out the common feature of the 10 pair correlations
shown in Figs.~\ref{fig:fig1-gdr1} and \ref{fig:fig2-gdr2} that the
most pronounced $T-$dependence is observed in the first peak in that it
sharpens when temperature is lowered.  This reflects the changes in
the local bonding of the structural units SiO$_n$ and BO$_n$ of the
silicate and borate sub-networks (discussed below), as well as in their
mutual connectivity which increases when $T$ decreases. Concerning  the
position of the maximum of the first peak, we find basically no change
for the Si-O and O-O pairs, located at 1.63 ${\rm \AA}$ and 2.62 ${\rm
\AA}$ respectively (see Figs.~\ref{fig:fig1-gdr1}a and d), and this is
consistent with the high concentration of SiO$_4$ tetrahedra which is more
than 60\% even at 4500~K (see Fig. \ref{fig:fig3-coord}). However,
one should recall that the present simulations have been carried out
at fixed density, which, due to the presence of the strong covalent
bonding characterizing the SiO$_4$ tetrahedron, makes it difficult to
change the local structure in a significant manner.

In contrast to this, the location of the first peak in the B-O and Na-O
pairs shifts to higher values  (see Fig.~\ref{fig:fig1-gdr1}c and d): For
the B-O pair, the first peak is located at 1.34 ${\rm \AA}$ at 4500~K,
while at 2200~K it is at 1.38 ${\rm \AA}$. This shift is related to the
changes in the B-O coordination, since with decreasing $T$ the percentage
of tetrahedral BO$_4$ units increases, and the mean B-O nearest-neighbor
distance is larger in tetrahedral BO$_4$ coordination than in trigonal
BO$_3$  coordination or in defect  BO$_2$  coordination present at
the highest temperatures (see next subsection). For the Na-O pair,
the first peak position shifts from  2.17 ${\rm \AA}$ at 4500~K to
2.24 ${\rm \AA}$ at 2200~K, and this is consistent with the increasing
network polymerization as it can be derived from the $T-$dependence of
the fractions of oxygen species found in our liquids: Bridging oxygens
(BO), non-bridging oxygens (NBO) as well as tricluster oxygens (TBO)
(see next subsection).  We recall that \bo~are oxygen atoms connected
to two network cations (Si and B), the \nbo are connected to only one
network cation, while the defective \tbo units are connected to three
network cations.

In Fig.~\ref{fig:fig2-gdr2}, we show the PDFs of the two network formers,
Si and B, as well as their correlations with sodium atoms, and the Na-Na
pair correlation. Cooling from 4500~K to 2200~K has the usual effect
that the structural order at short and intermediate distances (i.e. for
$r\le 7-8$ \AA) increases, i.e. the peaks and the minima become more
pronounced. In particular we note that in this $T-$range the height of
the first-nearest-neighbour peak changes by only $15-20\%$ for the sodium
correlations (Figs. \ref{fig:fig2-gdr2}d, e, and f). If the temperature
is lowered to 300~K, the resulting change in the PDF is more pronounced
which shows that the Na atoms settle into their preferred local structure
only at relatively low temperatures, in agreement with the high diffusion
constant found for this species (see below). (The same effect is observed
in the Na-O  correlation, see Fig.~\ref{fig:fig1-gdr1}c.)

Figure~\ref{fig:fig2-gdr2}b shows that the first peak in the
B-B correlation splits into two if $T$ is lowered from 2200~K to
room temperature with the first peak is located at around 2~\AA.
This effect is related to the high quench rate used in the simulations
which freezes defective BO$_n$ coordination polyhedra sharing also
edges and not only corners. This B--B peak at small distances is also
related to the presence of  a sharp peak  around 90$^{\mathrm o}$
in the B--O--B bond angle distribution (see Fig.~9 below), since if
one considers  the average B-O distances in trigonal and tetrahedral
borons (1.37~\AA\, and 1.47\AA, respectively) the resulting  B--B
distance is equal to 2~\AA.  Finally we note in panels a) and b)
of Fig.~\ref{fig:fig1-gdr1} that for the glass the second nearest
neighbor peak has several smaller peaks.  The latter ones reflect the
presence of SiO$_n$ and BO$_n$ coordination polyhedra ($n=3,4,5,6 $)
sharing not only corners but edges as well. Although the concentration
of  these defects decreases when temperature decreases (see below, and
also Refs.~\cite{Benoit2001,Du2006114702,Ispas2002,Pohlmann2004184209}),
the involved distances become better defined and hence the corresponding
peaks show up in the PDF only at low temperatures.

\subsection{\label{sec:coordliq} Coordination Numbers}

\begin{figure}
\includegraphics[width=0.43\textwidth]{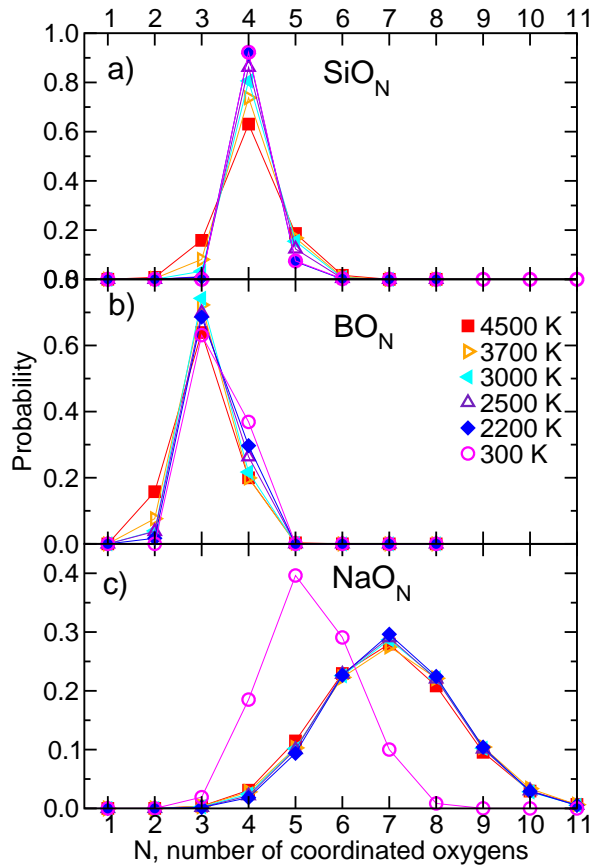}%
\caption{\label{fig:fig3-coord} Distributions of Si-O, Na-O and B-O
coordinations for the liquid and glass at 300~K.  The temperatures for the liquid are:
4500K, 3700K, 3000 K, 2500 K, and 2200 K. }
\end{figure}

A more detailed characterization of the local structure is given by
the distribution of the coordination numbers.  These distributions have been 
computed by defining two atoms to be nearest neighbours if their distance
is less than the position of the first minimum of the corresponding
PDF $g_{\alpha\beta}(r)$. Since we have found that this position is
basically independent of temperature for the liquid, we have used the
values 2.35 \AA, 2.15 \AA\,, and 3.40 \AA\, for the Si-O, B-O and Na-O
bond distance, respectively. For the glass state, we have used 2.0 \AA\,
for both  Si-O and B-O bond distance, while a cutoff equal to 3.0 \AA\,
has been used for the Na-O  coordination.  Figure \ref{fig:fig3-coord}
shows the distributions of the oxygen coordinations with Si, Bi, and Na
atoms for the five liquid temperatures as well as for the glass state.


For the Si-O coordination, shown in Fig.~\ref{fig:fig3-coord}a, we find
the expected trend, already reported for liquid silica or more complex
silicate liquids \cite{Benoit2001,Ispas2002}, that with decreasing
temperature  the percentages of defect units SiO$_3$ and SiO$_5$
decreases in favor of increasing of SiO$_4$ tetrahedral units. Hence at
the highest temperature, one has 60${\rm \%}$ of Si atoms in a SiO$_4$
tetrahedral unit, while at 2200~K this percentage reaches 95${\rm \%}$.
We also note  that, below 2500~K, there are no more under-coordinated
SiO$_3$ units, and that, at the lowest liquid temperatures, the fraction
of over-coordinated  SiO$_5$ units is around $5\%$ .

\begin{figure}
\includegraphics[width=0.43\textwidth]{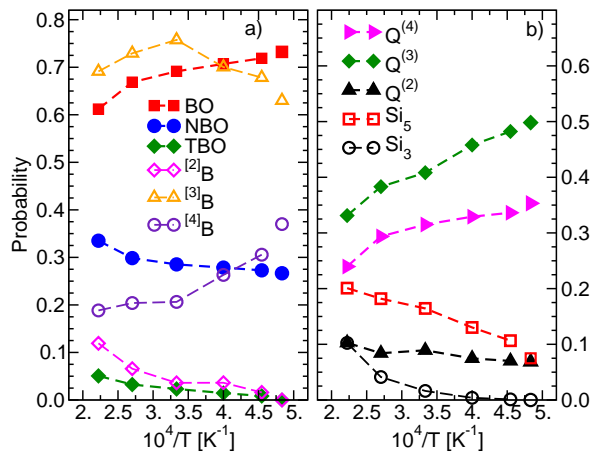}
\caption{\label{fig:fig4-oxygen} Temperature dependence of the
concentrations of BO, NBO and tricluster oxygen TBO, $^{[n]}$B for
$n=2,3,4$  (panel a), and of the $Q^{(n)}$ units for $n=1,2,3,4$
where $Q^{(n)}$ is a SiO$_4$ tetrahedron with $n$ bridging oxygens
for $n=1,2,3,4$, as well as of the defective units Si$_5$ and Si$_3$
(panel b). The symbols that are not connected by the lines correspond
to the values for the glass state. Note that the two panels do not have
the same scale on the ordinate.
}
\end{figure}

The distribution for the B-O coordination, shown  in
Fig.~\ref{fig:fig3-coord}b, indicates that, at the highest $T$,
one has around  $20\%$ BO$_2$ units and that cooling causes the
gradual disappearance of this local structure in favor of formation
of trigonal and tetrahedral borons. Since in the $T-$range 4500~K to
3000~K the concentration of BO$_4$ units is basically constant we can
conclude that BO$_2$ is mainly converted into BO$_3$ units (see also
Fig.~\ref{fig:fig4-oxygen}a). For $T\leq 3000$~K the concentration of
BO$_2$ units is less than $3\%$, and we note the conversion of trigonal
borons to tetrahedral ones and at 2200 K the borate sub-network is
essentially made up of BO$_3$ ($\approx 69 \%$) and BO$_4$ ($\approx 29
\%$)  units. These changes in boron coordination with temperature are in
qualitative agreement with the results of high temperature NMR experiments
\cite{Sen199829,Sen2007094203,Stebbins200880,Wu20102097,Wu20113944} for
alkali borosilicates and boro-aluminosilicates. From these experiments
it was concluded that the reaction BO$_3 +$NBO $\rightarrow$ BO$_4$
takes place.  However, if this would indeed be the main mechanism
responsible for the  conversion of BO$_3$ units into BO$_4$, the NBO
concentration should decrease in the same manner as the increase of the
BO$_4$ concentration, especially between 3000~K and 2200~K. Since this is
not what happens for our system, as can be seen in Fig.~\ref{fig:fig4-oxygen}a
discussed below: The transformation of BO$_3$ units into BO$_4$ is
more complicated than the above mentioned speciation reaction in that
part of the transformation must involve a simultaneous change of the
silicate sub-network.

The temperature dependence of the Na coordination number, shown in
Fig.~\ref{fig:fig3-coord}c, is somewhat surprising. For all the liquid
states considered the distribution is basically independent of $T$,
with most Na atoms having 6-8 neighbors. If, however, the samples are
cooled to the glass state, this distribution shows a strong shift in
that the new maximum is located at 5, with most Na atoms having between
4 and 7 neighbors. This very strong change is coherent with the results
from the PDFs shown in Figs.~\ref{fig:fig1-gdr1} and \ref{fig:fig2-gdr2}
for which we found that these functions involving Na showed only a mild
$T-$dependence in the liquid state but then changed quickly if the system
is quenched into the glassy state.

Figure  \ref{fig:fig4-oxygen}a shows  the temperature dependence of
the oxygen speciations (BO, NBO, and TBO, filled symbols), together
with that of the boron units $^{[i]}$B, for $i=2,3,4$, where $i$ is the
number of oxygen neighbors (open symbols). At the highest temperature,
the TBO concentration is around 5\%, and then decreases rapidly following
basically an Arrhenius law, so that this species has almost disappeared
at the lowest liquid temperature. From the figure we can also conclude
that with decreasing temperature the network connectivity increases
since the BO\, concentration increases quite quickly whereas the NBOs
fraction is decreasing (and both $T-$dependencies are compatible with
an Arrhenius law).  If this $T-$dependence is extrapolated to 760~K,
i.e. the experimental $T_g$, one finds a concentration of around 90\% for
BO and 10\% for NBO, respectively.  These values are not too far from the
ones predicted by the YBD model (80\% for BO and 20\% for NBO)
and also compatible with estimates from experiments~\cite{Wang19991519}.


Regarding the temperature dependence of the boron coordinations, the data
plotted in Fig. \ref{fig:fig4-oxygen}a support the scenario mentioned
above that $^{[2]}$B units are transformed into trigonal ones at the
highest temperatures, since $^{[4]}$B is basically independent of $T$,
followed by the conversion of $^{[4]}$B into tetrahedral units at lower
temperatures. This reaction if probably only a first order approximation,
since one can expect that there is also an interplay between the above
mentioned changes and the ones related to the silicate sub-network,
shown in Fig.~\ref{fig:fig4-oxygen}b. Although in the temperature range
in which we can equilibrate the liquid the majority of boron atoms is
3-fold coordinated, the $T-$dependence shown in Fig.~\ref{fig:fig4-oxygen}a
shows that this concentration is decreasing rapidly and that the one of
$^{[4]}B$ is increasing. If one makes a reasonable extrapolation of
this trend to the experimental glass transition temperature one predicts
that at $T_g$ the concentration of $^{[4]}$B  is around 75\% (and 25\%
$^{[3]}$B). For a more detailed discussion see Fig.~2 in Part~II as well
as the accompanying text.

For the silicate sub-network, a complementary information on its
connectivity is given by the $T-$dependence of the Q$^{(i)}$ species,
for $i=2,3,4$, plotted in Fig.  \ref{fig:fig4-oxygen}b, together with
the percentages of 3-and 5-fold coordinated silicon atoms, denoted by
Si$_3$ and Si$_5$, respectively. (We recall that  Q$^{(i)}$ is a SiO$_4$
tetrahedron with exactly $i$ bridging oxygens, and since there are almost
no  \Q1 units nor 6-fold coordinated Si, we do not show them.)

Firstly we note that the Si$_3$ concentration follows quite closely an
Arrhenius law for decreasing temperatures and becomes basically zero at
the lowest temperatures we have studied the liquid. Also the concentration
of the Si$_5$ units decreases quite rapidly and a simple extrapolation
to the experimental $T_g \approx 760$~K shows that at this temperature
its concentration is also very close to zero. The $T-$dependence of the
Q$^{(i)}$ species shows that the concentration of Q$^{(3)}$ increases
rapidly, the one for Q$^{(4)}$ a bit slower, and the one for Q$^{(2)}$
decreases. In fact we find that the $T-$dependence of Q$^{(4)}$ cancels
the one of Si$_3$ to a high accuracy (1\% level) which suggests that if,
with decreasing $T$, a 3-fold coordinated Si atom picks up an oxygen
neighbor, it transforms into a Q$^{(4)}$ unit, i.e. it has no dangling
oxygens.  On the other hand, if a 5-fold coordinated Si atom sheds one
of its oxygen neighbors, it will transform into a Q$^{(3)}$ unit (and to a
smaller extent into a Q$^{(2)}$ unit). (This can be inferred from the fact
that the sum of concentrations Si$_3$, Q$^{(3)}$, and Q$^{(2)}$ is only
very weakly $T-$dependent.) This result is reasonable since it can be
expected that the Si$_5$ unit was locally negatively charged and when
it lost one oxygen it became a bit positively charged, thus impeding
that the remaining oxygen atoms from bridging bonds.  The oxygen freed
by the Si$_5$ becomes now available to transform a  \biii\,
 unit into a  \biv. Thus we see that with this scenario the conversion of
\biii\,  into \biv\, is intimately linked to the $T$-dependence of the
structure of the silicate sub-network.

\subsection{\label{sec:Structfactor}Structure factors }

\begin{figure}
\includegraphics[width=0.43\textwidth]{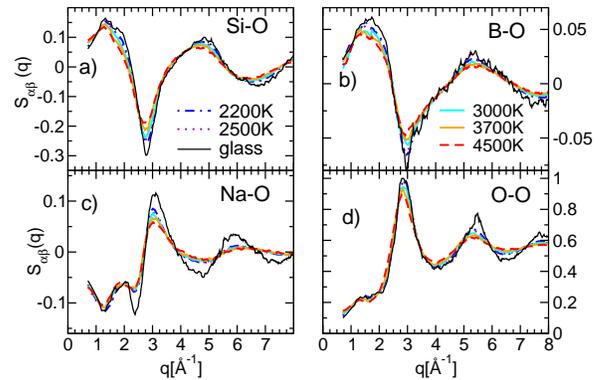}%
\caption{\label{fig:fig5-sq1} Partial structure factors for liquid
and glassy NBS, of X-O pairs, for X=Si, B, Na, and O. Note that  the
different panels do not have the same scale on the ordinate.}
\end{figure}

\begin{figure}
\includegraphics[width=0.40\textwidth]{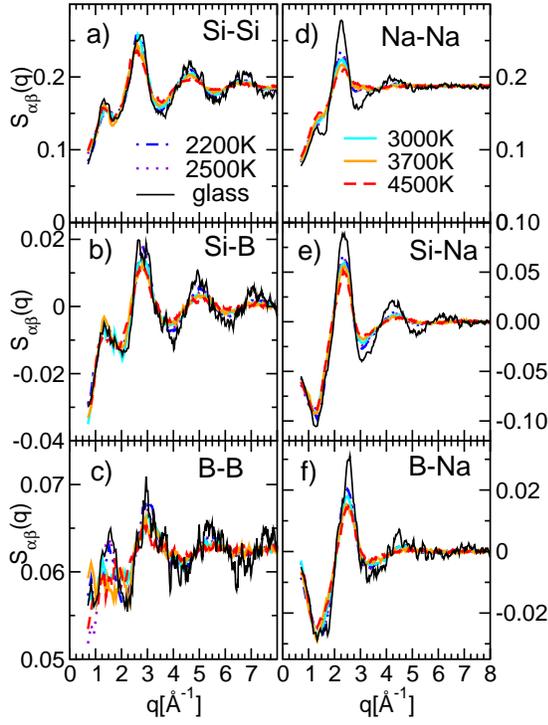}
\caption{\label{fig:fig6-sq2} Partial structure factors
$S_{\alpha\beta}(q) $  for $\alpha, \beta =$ Si, B, Na.  Note that
the different panels do not have the same scale on the ordinate. }
\end{figure}

The pair distribution functions are useful quantities to characterize the
structure of a liquid at short distances. However, for intermediate and
long distances it is better to consider their space Fourier transform,
i.e. the partial structure factors. Figures~\ref{fig:fig5-sq1} and
\ref{fig:fig6-sq2} show the 10 partial structure factors characterising
our liquid and glass, where $S_{\alpha\beta}(q)$ has been computed using
the definition \cite{JHansen_book,BinderKob_book}:

\begin{equation}
S_{\alpha\beta}(q) =  \frac{f_{\alpha\beta}}{N} 
\sum_{j=1}^{N_\alpha} \sum_{k=1}^{N_\beta} 
\left \langle \exp(i{\bf q}.({\bf r}_j - {\bf r}_k)) \right \rangle
\quad  \alpha,\, \beta = {\mathrm{Si, O, Na, B} }, 
\end{equation}

\noindent
Here f$_{\alpha\beta}=1$ for ${\rm \alpha=\beta}$ and
f$_{\alpha\beta}=1/2$ otherwise and $N$ is the  total number of atoms.

As common features, we can firstly notice that each of the partial
factors shows either a main peak or a negative dip for $q$ vectors around
2.5-3.3\, $\mathrm \AA^{-1}$, which reflects the local bonding inside the
local structural units SiO$_n$ and BO$_n$. The second characteristics
is the presence of a less pronounced peak or negative dip at smaller
wavevectors around $1.2-1.4\, \mathrm \AA^{-1}$, which is related to
the so-called first sharp diffraction peak (FSDP) and corresponds
to the length scale associated to two connected SiO$_4$ tetrahedra
\cite{Horbach19993169,Horbach2008244118} and/or two trigonal boron
units \cite{Ohmura2008224206}. We note that the FSDP for the O-O pair
is only a bump (see Fig.~\ref{fig:fig5-sq1}d), which indicates that the
distribution of distances between two SiO$_4$ tetrahedra is broadened
since the oxygen atoms can be connected either to Si or B atoms.
For pure boron oxide liquid at 2500~K the {\it ab initio} molecular
dynamics simulations by Ohmura and Shimojo \cite{Ohmura2008224206}
have shown the existence of pronounced peaks or a negative dip at $1.6\,
\mathrm \AA^{-1}$, $2.4\, \mathrm \AA^{-1}$ and $3.0\, \mathrm \AA^{-1}$
for the three pair correlations B-O, O-O and B-B, and our corresponding
NBS data present similar features, although slightly shifted due to
the additional presence of Si and Na atoms. However, in our case the
B-B correlation, Fig.~\ref{fig:fig6-sq2}c, these features are not very
pronounced (note the scale in the graph) since in our composition the
concentration of B$_2$O$_3$ is relatively low. Finally we point out that
the Si-B correlation shown in Fig.~\ref{fig:fig6-sq2}b does not seem to
go to zero for $q\to 0$ in the accessible $q$-range, which indicates the
presence of a microphase separation of the two sub-networks. Previous
classical MD simulations have indeed mentioned the tendency that
the borate network separates from the silicate part together with
an association of sodium with the former~\cite{Gou2001539}. But this
conclusion was rather qualitative as it was based only on snapshots of
the system. Instead here we find that this trend is indeed observable
in a correlation function.

Concerning the sodium correlations (see $S_{\mathrm{NaO}}$
in Fig.~\ref{fig:fig5-sq1}c, as well as $S_{\mathrm{NaNa}}$,
$S_{\mathrm{SiNa}}$, and  $S_{\mathrm{NaB}}$ in Fig.~\ref{fig:fig6-sq2}d,
e, and f respectively), we note that they show a characteristic feature
at about $1.3-1.4~{\rm \AA^{-1}}$ for all considered temperatures:
a shoulder for $S_{\mathrm{NaNa}}$, and a negative dip for the
three other correlations. In analogy to binary sodosilicate liquids
\cite{Greaves1985203,Horbach2002125502,Meyer2004027801}, we can interpret
this feature as indication for the presence of sodium rich channels in
the structure, thus concluding that such intermediate range structures
exist also in multicomponent glass-formers.

\begin{figure}[b]
\vspace*{4mm}
\includegraphics[width=0.43\textwidth]{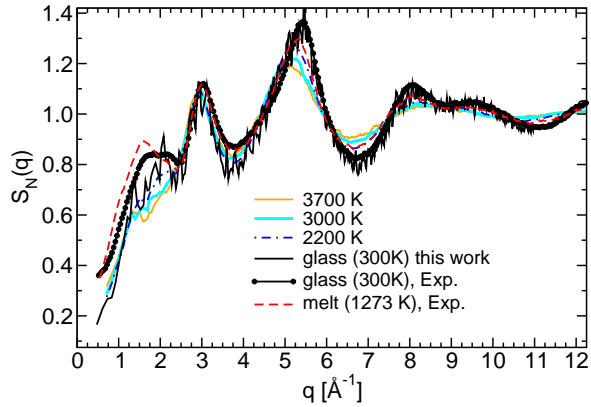}%
\caption{\label{fig:fig7-sqn} Calculated
and experimental neutron structure factor $S_N(q)$ for liquid and glass states.
The experimental curves are from Ref.~\protect \cite{Michel2013169} }
\end{figure}

>From the partial structure factors one obtains immediately the total
static structure factors which can be compared to  experimental results
if available. We first consider the neutron structure factor, which is a
linear combination of the partial structure factors \cite{BinderKob_book}:

\begin{equation}
S_{\mathrm N}(q)= \frac{N}{\sum_{\alpha=\mathrm{Si, O, B, Na}} N_\alpha b_\alpha^2}
\sum_{\alpha, \beta=\mathrm{Si, O, B, Na}}  b_\alpha b_\beta S_{\alpha \beta} (q)
\end{equation}

\noindent
with the neutron scattering length $b_\alpha$ given by
$b_{\mathrm{Si}}=4.1491$ fm, $b_{\mathrm O}=5.803$ fm, $b_{\mathrm
B}=6.65$ fm, and $b_{\mathrm{Na}}=3.63$ fm, respectively \cite{nist}. In
Fig.~\ref{fig:fig7-sqn} we present $S_{\mathrm N}(q)$ for the   liquid
and glass states together with recent experimental results for the melt at
1273~K and the glass~\cite{Michel2013169}. From Fig.~~\ref{fig:fig7-sqn}
we can conclude that in general there is a rather good agreement
between the results from the simulations and the experimental data.
Several features can be noted: Firstly, there is a main peak around
5.0~${\rm \AA}^{\rm -1}$ which becomes more pronounced and slightly
shifts to higher $q$ with decreasing temperature in both experimental
and simulation data.  Note that although this peak is the highest one in
$S_{\mathrm N}(q)$, it does not really correspond to a particular feature
in the partial structure factors, hence making the interpretation of
$S_{\mathrm N}(q)$ rather difficult if one does not have access to the
partials structure factors.  Secondly both experimental and simulation
data present a  peak around 2.8-3.0~\AA$^{-1}$, with a slight lower
intensity for the simulation curves.  This peak originates from the
peaks present in almost all the partial structure factors in the same
$q-$range and reflects the local bonding inside the structural units
SiO$_n$ and BO$_n$.  The peak position and intensity seem to be basically
independent of temperature   although the partial structure factors
do show a significant $T-$dependence (see Figs.~\ref{fig:fig5-sq1}
and \ref{fig:fig6-sq2}). Concerning the features at small $q$ (peaks
around 1~${\rm \AA}^{\rm -1}$ and 2~${\rm \AA}^{\rm -1}$) we see that
they are significantly less pronounced than in the partial factors shown
in Figs. \ref{fig:fig5-sq1} and \ref{fig:fig6-sq2}, thus indicating the
difficulty to observe them in a neutron scattering experiment.


A further quantity which can be obtained from the partial structure
factors and which is experimentally accessible is the  X-ray total
structure factor ${S_X}(q)$  which is given by \cite{Fischer2006233}:

\begin{equation}
S_{\rm X}(q)= \frac{N}{\sum_\alpha N_\alpha f^2_\alpha (q/4\pi)}
\sum_{\alpha, \beta}  f_\alpha (q/4\pi) f_\beta (q/4\pi)S_{\alpha\beta} (q)\, .
\end{equation}

\begin{figure}
\includegraphics[width=0.43\textwidth]{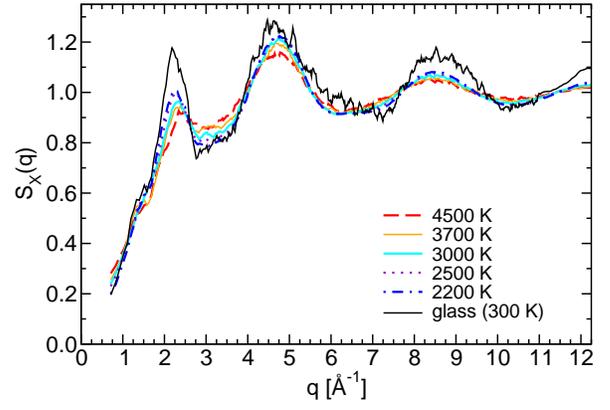}%
\caption{\label{fig:fig8-sqx} Calculated X-ray structure factor $S_X(q)$
for liquid and glass states. }
\end{figure}

\noindent
Here $f_\alpha (s)$ is the scattering-factor function (also called form
factor), computed as a linear combination of five Gaussians using the
parameters derived by Waasmaier and Kirfel \cite{Waasmaier1995416}. The $q-$
dependence of $S_{\mathrm X} (q)$ is shown in Fig.~\ref{fig:fig8-sqx}.
In contrast to the neutron total structure factor, the X-ray total
structure factor shows a pronounced peak around $2.1-2.3$ ${\rm
\AA}^{\rm -1}$ and then a second one around $4.5-4.7$ ${\rm \AA}^{\rm
-1}$.  When temperature is lowered, the first peak gains in intensity
and its position shifts to smaller $q$, while the second peak only
increases its intensity.  The shoulder seen in $S_{\mathrm N}(q)$ around
1.2--1.4~${\rm \AA}^{\rm -1}$ is hardly visible in $S_{\mathrm X} (q)$
so that one can conclude that, in experiments, the channel-like structure
can be detected more easily by means of neutron scattering instead of X-ray
scattering. As it is the case for S$_{\mathrm N}(q)$, we notice that the various
peaks in $S_{\mathrm X} (q)$ are difficult to interpret since they
are sum of too many partials. 


\subsection{\label{sec:BADliq} Bond Angle Distributions (BAD)}

\begin{figure}
\includegraphics[width=0.43\textwidth]{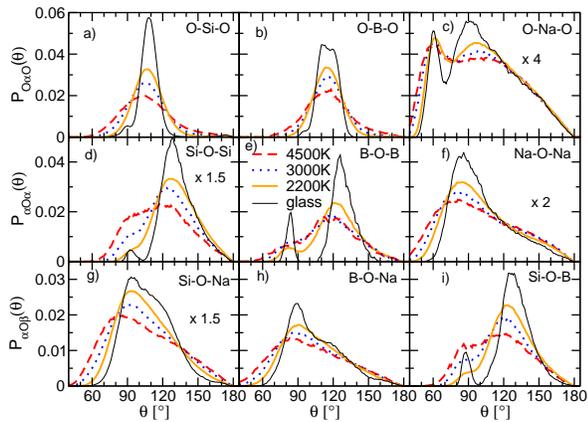}
\caption{\label{fig:fig9-bad} Angular distribution functions $P_{\mathrm
O\alpha \mathrm O} (\theta)$,  $P_{\alpha \mathrm O \alpha}(\theta)$,
and $P_{\alpha \mathrm O\beta} (\theta)$ for  ${\alpha,\, \beta,} =$  Si, B, and
Na shown for 3 temperatures in the liquid as well as for the glassy
state. Distributions in panels c), d), f) and g) have been multiplied
by the factor given in the panel in order to enhance their visibilities.}
\end{figure}

In order to get insight into the local connectivity within and between the
local building blocks of the network, we have determined the distribution
functions $P_{\alpha\beta\gamma}(\theta)$ of the bond angle formed by
the triplet $\alpha-\beta-\gamma$. Among all possible combinations of
triplets, we show in Fig.~\ref{fig:fig9-bad} the 3 sub-sets $\mathrm O -
\alpha-\mathrm O$,  $\alpha-\mathrm O - \alpha$ and  $\alpha-\mathrm O
- \beta$ for $\alpha,\, \beta=$ Si, B, Na. As usual one finds that the
distributions become more narrow if temperature is decreased.  For the
intra-tetrahedral angle OSiO, Fig.~\ref{fig:fig9-bad}a, the distribution
is quite broad at the highest temperature reflecting the presence of 3-,
5- and 6-fold coordinated Si, as well as distorted SiO$_4$ tetrahedra
with NBOs. If $T$ is lowered, we notice  a slight shift of its maximum
to higher angles and the formation of a small peak at 90$^{\mathrm o}$
which corresponds to the presence of 5-fold coordinated Si.

Regarding the OBO angle, i.e. the intra-coordination polyhedron of the
other network-former, Fig.~\ref{fig:fig9-bad}b shows that the distribution
is Gaussian-like for the liquid, but becomes split once the glass state
is reached. In Part~II \cite{Pedesseau-nbs2}, we will discuss the
decomposition of this distribution for the glass into a contribution
coming from the trigonal BO$_3$ units and tetrahedral BO$_4$ units.

The distribution $P_{\mathrm{ONaO}}$, Fig.~\ref{fig:fig9-bad}c, is very
broad, and one recognize two contributions: One at 60$^{\mathrm o}$, and
a second one quite asymmetric around $90^{\mathrm o}$ with a long tail
towards larger angles. For the NBS glass, we will discuss the origin
of this double peak shape, already reported in previous simulations
for low-silica alkali-alkaline earth melts \cite{Tilocca2010014701}
in Sec. II.B of the companion paper \cite{Pedesseau-nbs2}).


For the so-called inter-tetrahedral angle SiOSi, shown in
Fig.~\ref{fig:fig9-bad}d, we have at 4500~K a broad distribution
and when cooling to the lowest liquid temperature, its widths
reduce significantly.  At 2200~K, the function $P_{\mathrm{SiOSi}}$
shows a maximum around 130$^{\mathrm o}$, but also a shoulder around
90$^{\mathrm o}$. The latter is due to the presence of edge-sharing
tetrahedra, as has been seen in other simulations of more simple
silicates \cite{Benoit2001,Pohlmann2004184209,Du2006114702}.  The BOB
distribution, Fig.~\ref{fig:fig9-bad}e, shows almost no $T-$dependence
between 4500~K and 3000~K and is very broad, but once one has reached
2200~K one recognizes the presence of a shoulder at around 85$^{\mathrm
o}$, which is related to BO$_3$ and BO$_4$ units that share an edge,
and a pronounced peak around 125$^{\mathrm o}$. Note that at low $T$
the distribution for SiOSi and BOB are qualitatively similar, which
shows the similar role played by the network formers. However, we also
recognize that B leads to significantly more edge sharing units than Si.

The last distribution involving the connectivity between
two network-former units is the one for SiOB, shown in 
Fig.~\ref{fig:fig9-bad}i. One sees that it is qualitatively similar to
the distribution $P_{\mathrm{SiOSi}}$ and also the different peaks can
be interpreted in an analogous manner.

For the distribution $P_{\mathrm{NaONa}}$, shown in
Fig.~\ref{fig:fig9-bad}f, decreasing temperature affects both
its width and maximum position in that it narrows and shifts to
slightly higher angles, respectively. For the last two functions
$P_{\mathrm{SiONa}}$ and $P_{\mathrm{BONa}}$, see panels g and h
in Fig.~\ref{fig:fig9-bad}, we see that lowering $T$ leads to a
significant decrease of the probability at large angles. As it will be
discussed in Ref.~\cite{Pedesseau-nbs2} this decrease is related to the
fact that Na is  avoiding the direction of the Si-O bond (or B-O bond).

\section{\label{sec:DP}Dynamical properties }

In the previous subsections we have discussed the structural properties
of the liquid as a function of temperature. In the following we will
concentrate on the dynamical features of the system. In particular we will
present the mean squared displacement of a tagged particle, the van Hove
correlation function, as well as the intermediate scattering function and
discuss how these dynamical quantities are related to the structural ones.

\subsection{\label{sec:MSD}Mean squared displacement and diffusion constant}

The mean squared displacement (MSD) of a tagged particle of type
$\alpha$, $\alpha=\{\mbox{Na, O, B, Si}\}$ is given by
\cite{JHansen_book,BinderKob_book}:

\begin{equation}
r^2_\alpha (t) =\frac{1}{N_\alpha}\sum_{i=1}^{N_\alpha} \langle |\vec r_i (t) -\vec r_i(0)|^2\rangle .
\end{equation}

\begin{figure}
\includegraphics[width=0.43\textwidth]{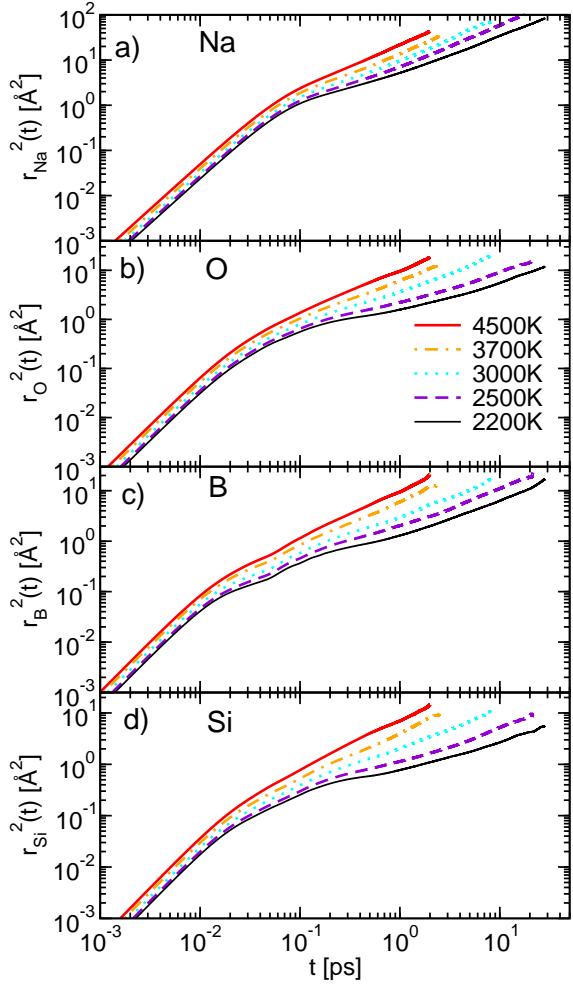}%
\caption{\label{fig:fig10-msd} Double logarithm plot of the mean squared displacement for Na
(a), O (b), B (c), and Si (d) atoms, versus time for the five temperatures
simulated.}
\end{figure}

In Fig.~\ref{fig:fig10-msd} we show the time dependence of the
MSD for the four atomic species in a double logarithmic plot. In
agreement with previous results for the MSD of glass-forming systems
\cite{Sarnthein199512690,Horbach19993169,Micoulaut2006031504,Hawlitzky2008285106,Horbach200187,Pohlmann2004184209,Du2006114702,Ganster200410172,Tilocca2007224202,Tilocca2010014701}
we find at short times a ballistic regime, i.e. $r^2_\alpha(t) \propto
t^2$, and at long times the diffusion behavior, $r^2_\alpha(t) \propto t$.
The fact that, for all temperatures considered, the MSD at long times
shows a diffusive behavior is evidence that the runs are sufficiently
long to fully equilibrate the system. Note that the time at which the
ballistic regime ends depends on the species and hence also the value of
the MSD that is reached at this crossover time depends on $\alpha$. This
shows that the mean free path for the particles, or the size of their cage
at low temperatures, does depend on the species. In particular we find
that for the Na atoms this size is significantly larger than the one
for the other species.

The curves for the sodium atoms show that the ballistic regime is almost
immediately followed by the diffusive regime which implies that even at
the lowest temperatures this species does not experience a significant
caging. This is in contrast to the behavior of the other types of atoms
for which one finds at intermediate time scales a relatively flat region
in the MSD, i.e. that the atoms are temporarily trapped. Furthermore
one sees that at low $T$ the curves for Si and B show in this caging
regime several shoulders. These are related to the rattling motion
of the atoms inside their cage which is  rather complex and involves
several frequencies and length scales (related to the position in time
and the value of the shoulders, respectively), in particular for the
B atoms. The details of this vibrational motion will be discussed in
Ref.~\cite{Pedesseau-nbs2} where we present the density of states of
the glass.


\begin{figure}
\includegraphics[width=0.43\textwidth]{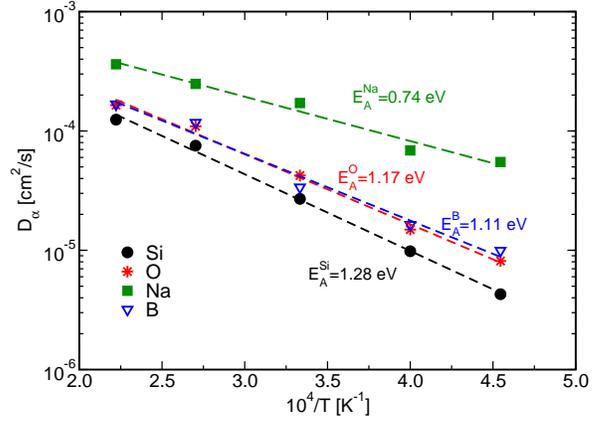}%
\caption{\label{fig:fig11-diffusion} Diffusion constants $D_\alpha$
for $\alpha=$ Na, O, B, and Si for the NBS liquid, plotted versus inverse
temperature.}
\end{figure}

>From the MSD at long times and the Einstein relation we can obtain the
self-diffusion constants:

\begin{equation}
D_\alpha=\lim_{t\to +\infty} \frac{r^2_\alpha (t)}{6t} \quad ,
\end{equation}

\noindent
and in Fig. \ref{fig:fig11-diffusion} we show the $T-$dependence of
$D_\alpha$ in an Arrhenius plot. This graph demonstrates that in the
$T-$regime considered the Na atoms are diffusing significantly faster
than the other species, i.e. their motion is decoupled from the one of
the other species, a result that is expected in view of the similar
behavior found in sodo-silicate melts~\cite{Horbach2002125502}. The
$T-$dependence is given by an Arrhenius law with an activation energy
around 0.74~eV. This value is only slightly smaller that the one
reported from classical MD simulations for a sodium disilicate having
a similar \soda\, concentration \cite{Horbach200187}, which shows that
these classical simulations give a fair estimate of the activation
energy. Electrical conductivity experiments for sodium borosilicates
that are rather boron rich have given, for temperatures above $T_g$, an
activation energy between 1.4 and 2.2~eV whereas viscosity measurements
give 0.65 and 0.85~eV~\cite{Ehrt2009185}. For temperatures below $T_g$, Wu
and coworkers \cite{Wu20112846,Wu20113661} have recently reported sodium
tracer diffusion results, and extracted activation energies between 0.71
and 0.83 eV. Thus on overall we can conclude that the results from our
simulation are compatible with the ones from experiments.


If we make the assumption that the Arrhenius law seen for Na holds also
for temperatures down to the experimental glass-transition, i.e.~760~K,
we can estimate that at $T_g$ the diffusion constant of the Na atoms
is around $8\cdot 10^{-8}$cm$^2$/s, which is close to the experimental
value reported for similar glass-formers~\cite{Wu20112846,Wu20113661}.
This shows that, although in our simulations we can access only relatively
high temperatures, it is possible to extract also useful information at
the experimental $T_g$.

In Fig.~\ref{fig:fig10-msd} we have seen that the motion for the Si atoms is the
slowest one. The diffusion constant allows to make this statement more
quantitative and from Fig.~\ref{fig:fig11-diffusion} we recognize that the
corresponding diffusion constant is, at the lowest temperature considered,
more than an order of magnitude smaller that the one for the sodium
atoms. The $T-$dependence of $D_{\rm Si}$ is also given by an Arrhenius
law and the activation energy is around 1.3eV, i.e. significantly larger
than the one for the sodium atoms. We note, however, that this value for
$E_A^{\rm Si}$ is by about a factor of four lower than the one found in
pure silica~\cite{Mikkelsen01,Horbach19993169}, which shows that the strong
depolymerisation of the network does lead to much faster diffusion.

The diffusion constant for the oxygen atoms is a bit higher than the one
of the silicon atoms in that the prefactor and the activation energy of
the Arrhenius law are slightly larger. This result is in qualitative
agreement with the behavior found in other silicate liquids and is
related to the fact that for a diffusive step the oxygen has to break
only one bond whereas a silicon atom has to break several ones.

Whereas the $T-$dependence of the diffusion constants for Na, Si, and O
is not very surprising, the one for boron is: Although the role of this
element is to be a network former and hence its dynamics should be slow,
we find that $D_{\rm B}$ is within the numerical accuracy identical to
$D_{\rm O}$. Hence, despite the fact that a typical B atom is connected
by three or four bonds to the matrix, it is still able to diffusive
relatively quickly.

The result that boron diffuses faster than Si is in qualitative
agreement with experiments on magmatic melts \cite{Baker1992617}. In
these experiments the diffusion constant was determined from viscosity
measurements at temperatures above T$_g$ for a system that had a similar
composition as the one considered here.  However, in a different type of
experiment it has been reported that the activation energy for the viscosity
is around 1.76~eV, i.e.~significantly higher than the one obtained
here~\cite{CochainPhD}.  Whether this difference is real or just due to
the fact that diffusion constant and viscosity do not necessarily have
the same activation energy (due to the breakdown of the Stokes-Einstein
relation) remains open.

\subsection{\label{sec:vanHove}Van Hove correlation function }

A more detailed understanding of the relaxation dynamics can be obtained 
from the self part of the van Hove function which is defined as~\cite{JHansen_book}

\begin{equation}
G_{s}^{\alpha}(r,t) = \frac{1}{N_\alpha} \sum_{i=1}^{N_\alpha}  
\left \langle \delta(r- |\vec r_i(t) -\vec r_i(0)|) \right \rangle
\quad  \alpha \in {\mathrm{Si, O, Na, B} }. 
\end{equation}

\noindent
Thus $G_{s}^{\alpha}(r,t)$ is the probability that in the time interval
$t$ a particle of type $\alpha$ has moved a distance $r$.

\begin{figure}

 \includegraphics[width=0.43\textwidth]{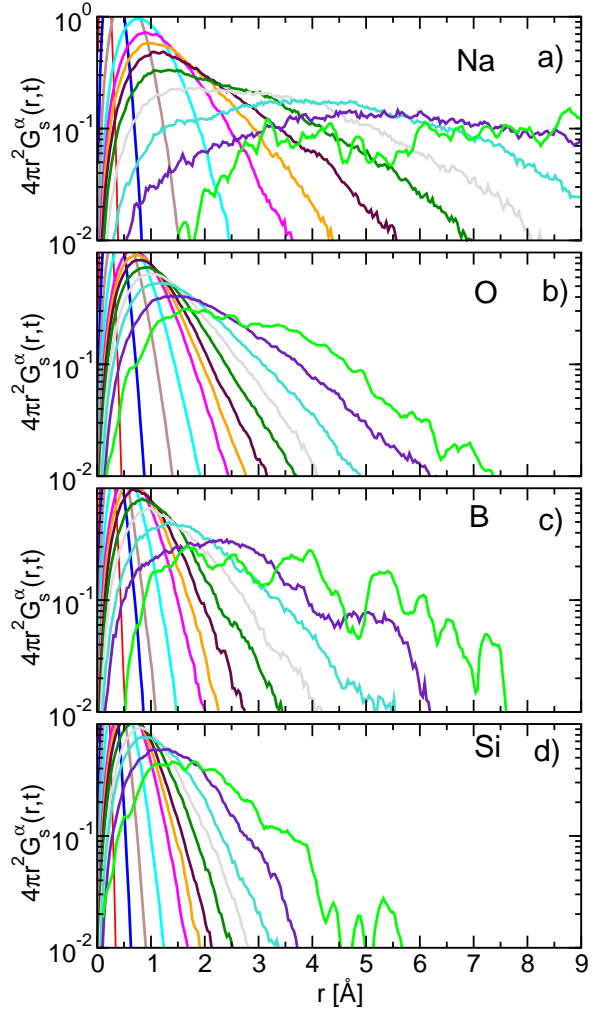}\ %
\caption{\label{fig:fig12-vanHove} The space dependence of the self-part
of the van Hove correlation function $G^\alpha_s (r,t)$  for $\alpha=$Na,
O, B, and Si, at 2200K and for different times.  The rightmost (green)
curve corresponds to $t=$30 ps, and then from right to left the times
are  $t=$ 15 ps, 7.5 ps, 3.75 ps, 1.9 ps, 0.9 ps, 0.45 ps, 0.225 ps,
0.1 ps, 0.05 ps, 0.025 ps and 0.0125 ps.
}
 \end{figure}

In Fig. \ref{fig:fig12-vanHove}, we show this function, multiplied by
the phase space factor $4\pi r^2$, for the four species at 2200 K, at
the following times:  0.0125ps, 0.025ps, 0.05ps, 0.1ps, 0.225ps, 0.45ps,
0.9ps, 1.9ps, 3.75ps, 7.5ps, 15ps, and 30ps.

Since at short times the atoms move ballistically, the self part of the van
Hove function is just a Gaussian, which explains the peak of the curves
seen in Fig.~\ref{fig:fig12-vanHove} for small $t$. For long times the
particles diffuse and hence the distribution of their displacements is
again a Gaussian. Thus the interesting information that can be obtained
from $G_s(r,t)$ are the deviations from this Gaussian behavior. For high
temperatures this Gaussian behavior is basically seen at all times and
hence we focus here on the lowest temperature for which some deviations
can be observed.

We see in Fig.\ref{fig:fig12-vanHove}a that for the sodium atoms the
distribution for $t=1.9$ps and $t=3.75$ps shows a weak shoulder at around
$r=3$. Since this distance corresponds to the nearest neighbor distance
between two Na atoms (see Fig.~\ref{fig:fig2-gdr2}f), we can conclude
that on this time scale there is an increased probability (with respect
to a purely diffusive process) that the atom which at $t=0$ was at the
origin has moved to this nearest neighbor distance. Such a behavior
is the signature of a hopping-like motion, a type of movement which
has been documented in previous classical simulations of sodo-silicate
systems~\cite{Horbach200187}  but so far not within {\it ab initio} simulations.


For the oxygen and boron atoms we see that, at a given $t$, the
distribution are more narrow than the one for the Na atoms, in agreement
with the observation that the diffusion constants of O and B are smaller
than the one for Na. For short and intermediate times the distributions do
not show any particular feature. However, for the longest times one can
notice a weak shoulder in $G_s(r,t)$ for oxygen at a distance around 3~\AA,
which is close to the nearest neighbor distance between two oxygen
atoms. Also for boron one see a peak at around 2.5~\AA~and a second
one at around 5~\AA, i.e. the distances corresponding to the first and
second nearest neighbor in the B-B correlation. Hence we can conclude
that also boron has the tendency to make a hopping-like motion.

For silicon the distributions are the most narrow ones (see
Fig.~\ref{fig:fig12-vanHove}d), in agreement with the fact that the
diffusion constant for silicon is the smallest one. We see that at the
two largest times also this distribution shows a small shoulder at round
$r=3$~\AA, i.e. the nearest neighbor distance between two Si atoms.

Finally we mention that if one plots $G_s(r,t)$, i.e. without the phase
space factor $4\pi r^2$, one finds that at intermediate times, 0.5~ps
$\leq t \leq 2$~ps, and distances $r\geq 1$~\AA~the distributions are
described well by an exponential law (not shown). In the past such a
behavior has been found also in other glass-forming systems (although
less complex ones) and it has been argued that this feature is related
to the fact that for short times the hopping motion of the particles do
not yet follow the central limit theorem~\cite{Chaudhuri2007060604}. From
our results we thus can conclude that the same mechanism is at work also
in this rather complex glass-former.

\subsection{\label{sec:scattering_function}Self intermediate function}

The dynamical quantities we have discussed so far, the MSD and $G_s(r,t)$,
are defined in real space. Although this makes the interpretation of
the observables easy, they are not accessible in a real experiment of
atomic systems, since scattering techniques probe the dynamics of the
system in reciprocal space. It is therefore important to understand
how the relaxation dynamics of our system would be seen in a scattering
experiment. Furthermore we have so far discussed only the time dependence
of single particles observables, which does not allow to make any
conclusion on the nature of the collective relaxation motion. In order to
address these points we will in the following discuss the time dependence
of the coherent and incoherent intermediate scattering functions.

The incoherent intermediate scattering function $F_s(q,T)$ is
defined as~\cite{JHansen_book}

\begin{equation}
F_s^{\alpha}(q,t) = N_{\alpha}^{-1} 
\sum_{j=1}^{N_\alpha} \exp [ i\vec{q} \cdot (\vec{r}_j(t)-\vec{r}_j(0))] \quad .
\label{eq12}
\end{equation}

\begin{figure}

\includegraphics[width=0.43\textwidth]{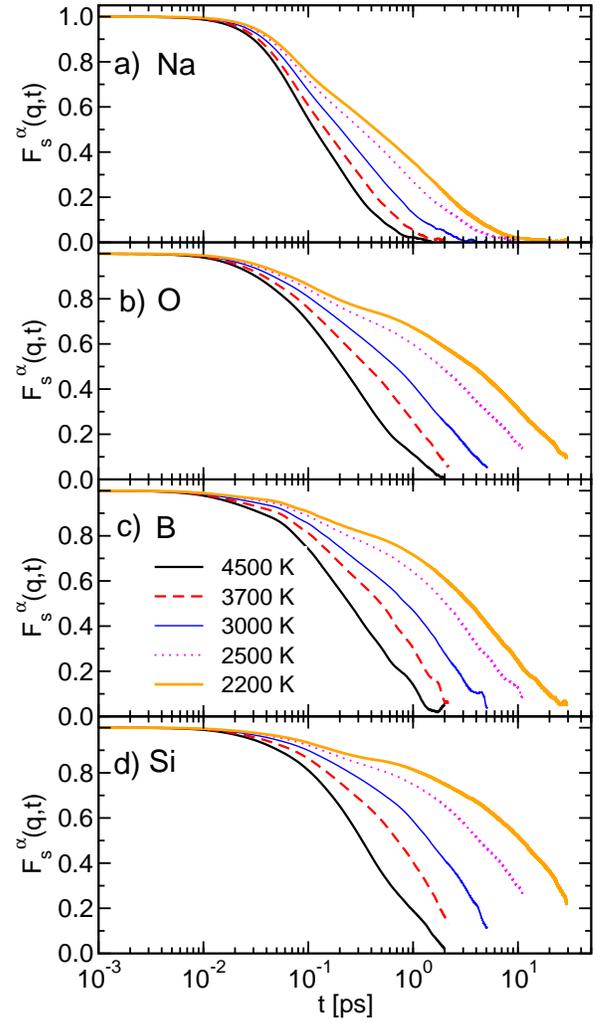}
\caption{\label{fig:fig13_fsqt} The time dependence of the incoherent intermediate scattering function
$F^\alpha_s(q,t)$ for $\alpha=$Na,
O, B, and Si, for the five temperatures simulated. The
wave-vector has been averaged over the range 1.05~\AA$^{-1}\leq q \leq$1.45~\AA$^{-1}$  in
order to improve the statistics.
}
\end{figure}

\noindent
Here $\vec{q}$ is the wave-vector and $q$ its modulus. In
Fig.~\ref{fig:fig13_fsqt}  we show the time dependence of $F_s(q,t)$ and in
order to improve the statistics  we have averaged the correlator over
wave-vectors in the range 1.05~\AA$\leq q \leq$1.45~\AA. Thus these
$q-$values are the ones for which we have in the partial structure
factors a pre-peak which is related to the presence of the channel-like
structure of the Na atoms (see Fig.~\ref{fig:fig7-sqn}), i.e we are looking
at the dynamics on length scales around 6~\AA. We mention, however,
that qualitatively similar results have been obtained also for other
wave-vectors. as long as $q$ is relatively small.

Figure~\ref{fig:fig13_fsqt}  shows that at high $T$'s the relaxation
is quick and that the correlators show basically an exponential
decay with relaxation times that are essentially independent of the
species. In contrast to this the correlators show at low temperatures
a two step decay, i.e. one sees at intermediate times a shoulder. This
feature is directly related to the relaxation dynamics inside the
cage~\cite{BinderKob_book}. We point out that also the correlator for Na
does show a weak shoulder, thus giving evidence that even this species
is somewhat caged, at least on this length scale. Although at low
$T$ the curves for the different species look qualitatively similar,
we recognize that the $\alpha-$relaxation times are very different,
in agreement with the strong species-dependence of the diffusion constant.


\begin{figure}

\includegraphics[width=0.43\textwidth]{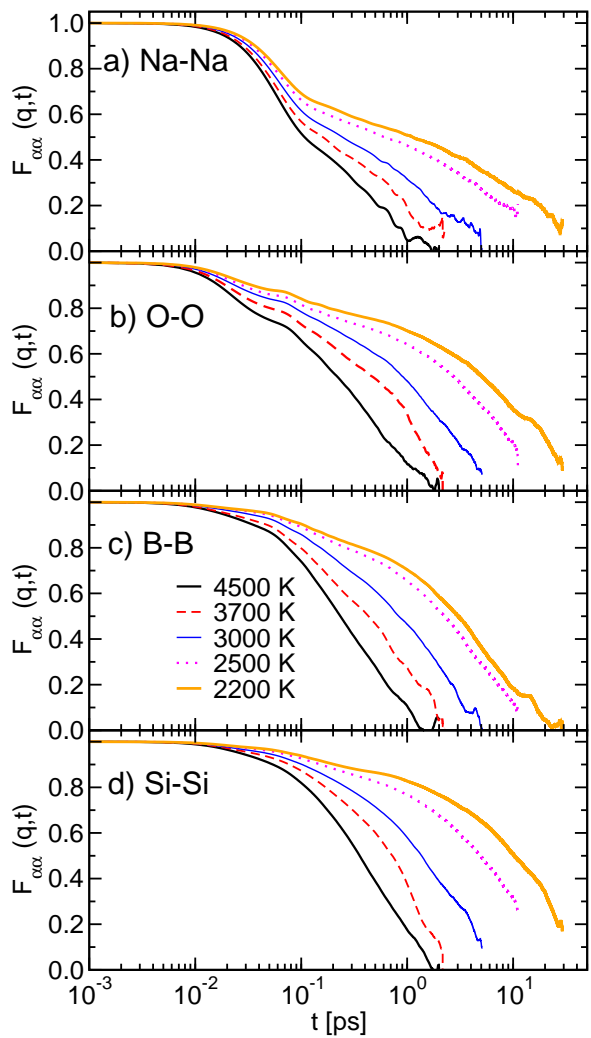}
\caption{\label{fig:fig14_fqt} The time dependence of the coherent
intermediate scattering function $F_{\alpha\alpha}(q,t)$ for $\alpha=$Na,
O, B, and Si, for the five temperatures simulated.  The
wave-vector has been averaged over the range 1.05~\AA$^{-1}\leq q \leq$1.45~\AA$^{-1}$  in
order to improve the statistics.
}
\end{figure}

In Fig.~\ref{fig:fig14_fqt}  we show the time  dependence of the coherent
intermediate scattering function $F(q,t)$ defined as~\cite{JHansen_book}

\begin{equation}
F_{\alpha\beta}(q,t) = f_{\alpha\beta} N_{\alpha}^{-1} 
\sum_{j=1}^{N_\alpha} \sum_{l=1}^{N_\beta} 
\exp [ i\vec{q} \cdot (\vec{r}_j(t)-\vec{r}_l(0)) ] \quad .
 \label{eq13}
\end{equation}

\noindent
Here the factor $f_{\alpha\beta}$ is 0.5 for $\alpha\neq \beta$
and 1.0 for $\alpha=\beta$. A comparison of the curves in
Fig.~\ref{fig:fig13_fsqt} with the one in Fig.~\ref{fig:fig14_fqt}
shows that for Si, O, and B the correlators are very similar in that
their shape and relaxation time are basically the same, and this holds
for all $T$.  This is the usual behavior found in glass-forming liquids
in that typically the self and collective functions decay on the same
time scale. We see, however, that the sodium atoms do not follow this
trend at all. Whereas the self function decays rather quickly, the
collective function decays on the same time scale as the one for the
other species. This clearly shows that in this system the Na atoms do
have sites that are favorable: Even if an atom stays only for a short
time at a given site (as can be concludes from $F_s^{\rm Na}(q,t)$),
a vacated site is quickly occupied by another Na atom. The time scale
to decorrelate the position of these ``special sites'' are therefore
the ones it takes the matrix to rearrange, i.e. is given by the motion
of the Si and O atoms. This result is in agreement with the one found
in sodo-silicate glass-formers in which a strong decoupling of the
coherent and incoherent correlator of the network modifier has already
been documented \cite{Horbach2002125502}


\section{\label{sec:conclu}Summary}

In this study, we have carried out first principles simulations
for a ternary sodium borosilicate liquid and glass, with a chemical
composition that is close to the one of glass wool.
 We have investigated
the local and intermediate range order structural features in the
liquid state with particular focus on how boron is embedded in the
silicate network.  We have found that in the liquid state the radial
distribution functions, as well as the partial structure factors that
involve Na, show a significantly smaller $T-$dependence than the ones
of the network formers. Furthermore we find that the partial structure
factors show at around 1.2~\AA$^{-1}$ a pre-peak which we see as evidence
for the existence of channel like structures than have been observed
in sodo-silicate systems before. However, our calculated X-ray and
neutron structure factor shows that with these type of experiments it is
difficult to see this pre-peak (with a slightly better chance for neutron
scattering).  The function $S_{\rm SiB}(q)$ does not seem to go to zero
even at the smallest wave-vectors accessible in our simulations. This
implies that these two network formers undergo a microphase separation
on the scale of a few nanometers.

Special attention has been given to the boron coordination, found to be
both trigonal and tetrahedral as expected for this composition. We have
found that for the temperatures at which we have been able to equilibrate
the liquid 60\% of boron is threefold coordinated and 40\% is fourfold
coordinated. However, the $T-$dependence of these concentrations clearly
shows that at the experimental glass transition temperature one expects
this ratio to be very different, with the concentration of $^{[4]}$B
reaching 70\%.

Regarding the dynamics we have determined the mean squared displacements
for all type of atoms and from them the diffusion constants. For all
species we find that $D_\alpha$ is given by an Arrhenius law with
activation energies that differ by a factor of 2 between Na and Si and
which are in reasonable good agreement with experimental values determined
at significantly lower temperatures. Surprisingly we find that the
diffusion constant of boron is very similar to the one for oxygen (and
significantly higher than the one of silicon). Thus for this system one
of the network-formers is significantly more mobile than the other one.

The space and time dependence of the self van Hove functions shows
that the Na atoms move, at the lowest temperatures, in a hopping-like
manner. Although a bit less pronounced we find the same behavior for the
boron atoms whereas the silicon atoms show a relaxation dynamics that is
much more flow-like. Hence we see that these two network formers have
a relaxation dynamics that differs not only quantitatively from each
other but also qualitatively.

Finally we have also determined the time dependence of the coherent
intermediate scattering function. We find that the ones for Si, O,
and B are very similar to the one for the incoherent functions, the one
for Na is very different from $F_s^{\rm Na}(q,t)$. In particular we see
that the former decays significantly slower than the latter, which is
again evidence that individual sodium atoms are moving in a channel-like
structure which relaxes only very slowly, i.e. on the time scale of the
rearrangement of the matrix.

\acknowledgements
We thank D. R. Neuville and B. Hehlen for  stimulating discussions on
this work. Financial support from the Agence Nationale de la Recherche
under project POSTRE is acknowledged.  This work was performed using HPC
resources from GENCI (TGCC/CINES/IDRIS) (Grants x2010095045, x2011095045
and x2012095045), and also on the HPC@LR cluster, Montpellier, France.
W. Kob acknowledges support from the Institut Universitaire de France.

\bibliography{main.bib}

\clearpage

\end{document}